\documentclass[twocolumn,showpacs,preprintnumbers,amsmath,amssymb,prl]{revtex4}

\usepackage{graphicx}
\usepackage{dcolumn}
\usepackage{amssymb}

\newcommand{\be}{\begin{equation}}
\newcommand{\ee}{\end{equation}}
\newcommand{\bea}{\begin{eqnarray}}
\newcommand{\eea}{\end{eqnarray}}

\newcommand{\lcm}{\begin{changemargin}{-1.8cm}{0.25cm}}
\newcommand{\ecm}{\end{changemargin}}
\newcommand{\tbcm}{\begin{changemargin}{-0.8cm}{-0.25cm}}
\newcommand{\tecm}{\end{changemargin}}

\begin{document}
\def\v#1{{\bf #1}}


\title{FFLO States in Holographic Superconductors}

\author{James Alsup$^{\natural}$}
\author{Eleftherios Papantonopoulos$^{*}$}
\author{George Siopsis$^{\flat}$}

\affiliation{{$^{\natural}$}Computer Science, Engineering and
Physics Department, The University of Michigan-Flint, Flint, MI
48502-1907, USA\\
$^{*}$Department of Physics, National Technical University
of Athens, GR-15780 Athens, Greece\\
$^{\flat}$Department of Physics and Astronomy, The University of
Tennessee, Knoxville, TN 37996 - 1200, USA}

\begin{abstract}
We discuss the gravity dual of FFLO states in strongly  coupled
superconductors.  The gravitational theory utilizes two $U(1)$
gauge fields and a scalar field coupled to a charged AdS black
hole. The first gauge field couples with the scalar
sourcing a charge condensate below a critical temperature, and the
second gauge field incorporates a magnetic field that couples to
spin in the boundary theory.  The scalar is neutral under the
second gauge field. By turning on a magnetic interaction between
the second $U(1)$ field and the scalar, it is shown that, in the
high-field limit, an inhomogeneous solution possesses a higher
critical temperature than the homogeneous case, giving rise to
FFLO states close to zero temperature.

\end{abstract}

\pacs{11.25.Tq, 04.70.Bw, 71.45.Lr, 71.27.+a} \maketitle


The application of the AdS/CFT correspondence  to condensed matter
physics  has developed into one of the most productive topics of
string theory.  It has opened up a broad avenue to understanding
strongly coupled phenomena of condensed matter physics by studying
their weakly coupled gravity duals. The holographic principle has
been applied to the study of conventional and unconventional
superfluids and superconductors \cite{HartnollPRL101}, Fermi
liquids \cite{Bhattacharyya:2008jc} and quantum phase transitions
\cite{Cubrovic:2009ye}.

The gauge/gravity duality, apart from superconductivity, has also
been applied to more general systems, characterized by additional
ordered states, like charge and spin density waves (CDW and SDW).
The development of these states corresponds to the spontaneous
modulation of the electronic charge and spin density, below a
critical temperature T$_c$.

A holographic CDW model was discussed in \cite{Aperis:2010cd}
consisting of two  scalar fields interacting via an antisymmetric
field and two St$\ddot{u}$ckelberg fields with a $U(1)$ Maxwell
gauge field. It was shown that below a critical temperature the
Maxwell scalar potential is modulated, corresponding to an
unidirectional modulated charge density in the conformal field
theory. Introducing a  modulated chemical potential, it was shown
in \cite{Flauger:2010tv} that below a critical temperature
superconducting stripes develop. Properties of the striped
superconductors and back reaction effects were studied in
\cite{Hutasoit:2011rd,Ganguli:2012up}. Striped phases were also
found in electrically charged RN-AdS  black branes that involve
neutral pseudo-scalars \cite{Donos:2011bh}.

Modulated order parameters  also appear as competing phases with
normal superconducting  phases in superconductor-ferromagnetic
(S/F) systems. They originate in high-field superconductors where
a strong magnetic field, coupled to the spins of the conduction
electrons, gives rise to a separation of the Fermi surfaces
corresponding to electrons with opposite spins (for a review see
\cite{Casalbuoni:2003wh}). If the separation is too high, the
pairing is destroyed and there is a transition from the
superconducting state to the normal one (paramagnetic effect).
Fulde and Ferrell \cite{Fulde} and Larkin and Ovchinnikov
\cite{Larkin} showed independently, that a new state could be
formed, close to the transition line. This state, known as the
FFLO state,  has the feature of exhibiting an order parameter,
which is not a constant, but has a space variation.  The space
modulation arises because the electron pair has nonzero total
momentum, and it leads to the possibility of a nonuniform or
anisotropic ground state, breaking translational and rotational
symmetries.

Holographic superconductors in the presence of an external
magnetic field have been discussed in the literature. It was found
that for a non-zero external magnetic field, it is inconsistent to
have non-trivial spatially independent solutions on the boundary
leading to two classes of localized solutions: the droplet
\cite{Albash:2008eh} and vortex solutions with integer winding
number \cite{Albash:2009iq, Montull:2009fe, Maeda:2009vf}. An
analytical study on holographic superconductors in external
magnetic field was carried out in \cite{Ge:2010aa}. A holographic
superconducting model with unbalanced fermi mixtures at strong
coupling was discussed in \cite{Bigazzi:2011ak}. In the background
of a charged AdS black hole, two $U(1)$ gauge fields were
introduced, one  sourcing the charge condensate and the other,
with a chemical potential imbalance, acting effectively on spins.
The charge and spin transport properties of the model were studied
but  the phase diagram did not reveal the  occurrence of FFLO-like
inhomogeneous superconducting phases.

Qualitatively, the FFLO phase formation in S/F systems may be
described in the framework of the generalized Ginzburg-Landau
expansion. In the standard Ginzburg-Landau functional
$F=a|\psi|^2+\gamma|\overrightarrow{\nabla}\psi|^2+\frac{b}{2}|\psi|^4,$
where $\psi$ is the superconducting order parameter, the
coefficient $a$ vanishes at the transition temperature $T_c$. At
$T<T_c$, the coefficient $a$ is negative and the minimum of $F$
occurs for a uniform superconducting state with $|\psi|^2=-a/b$.
If we  consider the paramagnetic effect of the magnetic field, all
the coefficients in the $F$ functional will be proportional to the
magnetic field $B$. In this case  qualitatively new physics
emerges due to the fact that the coefficient $\gamma$  changes its
sign at a point in the $(B,T)$ phase diagram indicating that  the
minimum of the functional does not correspond to a uniform state,
and a spatial variation of the order parameter decreases the
energy of the system. To describe such a situation it is necessary
to add a higher order derivative term in the expansion of $F$:
$F_G=a(B,T)|\psi|^2+\gamma(B,T)|\overrightarrow{\nabla}\psi|^2+\frac{\eta(B,T)}{2}|\overrightarrow{\nabla}^2\psi|^2+\frac{b(B,T)}{2}|\psi|^4$
(for a detailed account see \cite{Buzdin:2005zz}).

In this letter, we put forward a  gravity dual of FFLO states in
strongly coupled superconductors. In a dyonic black hole
background we introduce two $U(1)$ gauge fields and a scalar
field. The first gauge field has a non-zero scalar potential term
which in the boundary theory through its coupling to the scalar
field is the source of the charge condensate. The second $U(1)$
gauge field corresponds to an effective magnetic field acting on
the spins in the boundary theory. The scalar field is neutral
under the second $U(1)$ gauge field. There is a critical
temperature below which the system undergoes a second-order phase
transition and the black hole acquires hair. The system  possesses
inhomogeneous solutions for the scalar field which however always
give a critical temperature lower than the homogeneous one,
therefore the homogeneous solutions are dominant.

Next we turn on an interaction term of the magnetic field to the
scalar field of the generalized Ginzburg-Landau gradient type (in
a covariant form). The scalar field equation is modified and the
resulting inhomogeneous solutions give a critical temperature
which is higher than the homogeneous solutions. We attribute this
behaviour of the system to the appearance of FFLO states. Note
that the appearance of the FFLO states is more pronounced in the zero temperature limit (as the
magnetic field of the second $U(1)$ gauge group increases).

Consider the action \bea S &=& \nonumber \\ & \int & d^4 x
\sqrt{-g} \left[ \frac{R + 6/L^2}{16\pi G} - \frac{1}{4} F_{AB}
F^{AB}  - \frac{1}{4} \mathcal{F}_{AB} \mathcal{F}^{AB}\right],
\nonumber
\\ \eea where $F_{AB} =
\partial_A A_B - \partial_B A_A$, $\mathcal{F}_{AB} =
\partial_A\mathcal{A}_B -\partial_B \mathcal{A}_A$ are the field
strengths of the $U(1)$ potentials $A_A$ and $\mathcal{A}_A$,
respectively. We set $L=8\pi G = 1$.

The Einstein-Maxwell equations admit a solution which is a
four-dimensional AdS black hole of two $U(1)$ charges, \be ds^2 =
\frac{1}{z^2} \left[ - h(z) dt^2 + \frac{dz^2}{h(z)} + dx^2 + dy^2
\right],
\label{eqmetric1}
\ee with the horizon radius set at
$z=1$.

The two sets of Maxwell equations admit solutions of the form,
respectively, \be\label{eq3} A_{t} = \mu \left( 1 - z \right) \ , \ \ A_z
=A_x=A_y =0, \ee and \be\label{eq4} \mathcal{A}_y = \mathcal{B} x \ , \ \
\mathcal{A}_t = \mathcal{A}_x = \mathcal{A}_z = 0, \ee with
corresponding field strengths having non-vanishing components for an
electric and a magnetic field in the $z$-direction, respectively,
\be\label{eq5} F_{tz} = - F_{zt} = \mu \ \ , \ \ \ \ \mathcal{F}_{xy} =
-\mathcal{F}_{yx} = \mathcal{B}. \ee Then from the Einstein
equations we obtain \be\label{eq6}  h(z) = 1 -\left( 1 +
\frac{\mathcal{B}^2+\mu^2}{4} \right) z^3 + \frac{\mathcal{B}^2 +
\mu^2}{4} z^4 ~.\ee
The Hawking temperature is \be\label{eq7} T = - \frac{h'(1)}{4\pi}
= \frac{3}{4\pi } \left[ 1 - \frac{\mathcal{B}^2 + \mu^2}{12}
\right]. \ee In the limit $\mu, \mathcal{B} \to 0$ we recover the
Schwarzschild black hole.

We now consider a scalar field $\phi$, of mass $m$, and $U(1)^2$ charge $(q,0)$,
with the action \be\label{eq8} S = \int d^4 x \sqrt{-g} \left[ |D_A \phi |^2 -
m^2 |\phi|^2 \right], \ee where $D_A = \partial_A + iqA_A$.

The asymptotic
behavior (as $z\to 0$) of the scalar field is \be\label{eq9} \phi \sim z^\Delta \ \ , \ \ \ \ \Delta (\Delta -3) = m^2~. \ee
For a given mass, there are, in general, two choices of $\Delta$,
\be \Delta = \Delta_\pm = \frac{3}{2} \pm \sqrt{\frac{9}{4} +m^2}~, \ee
leading to two distinct physical systems.

As we lower the temperature, an instability arises and the system undergoes a second-order phase transition with the black hole developing hair.
This occurs at a critical temparture $T_c$ which is found by solving the scalar wave equation in the above background,
\be\label{eq19} \partial_z^2 \phi + \left[ \frac{h'}{h} - \frac{2}{z}
\right] \partial_z \phi + \frac{1}{h} \nabla_2^2\phi - \frac{1}{h}
\left[ \frac{m^2}{z^2} - q^2 \frac{A_t^2}{h}  \right] \phi = 0, \ee
with the metric function $h$ given in \eqref{eq6} and the electrostatic potential $A_t$ in \eqref{eq3}.

Although the wave equation \eqref{eq19} possesses $(x,y)$-dependent solutions, the symmetric solution dominates and the hair that forms has no $(x,y)$ dependence. To see this, let us introduce $x$-dependence and consider a static scalar field of the form
\be\label{eq11}  \phi (z , x ,y) = \psi(z) e^{iQx} ~.
\ee
The wave equation becomes
\be\label{eqw} \psi''
+ \left[ \frac{h'}{h} - \frac{2}{z} \right] \psi' - \frac{Q^2}{h}
\psi - \frac{1}{h} \left[ \frac{m^2}{z^2} - q^2 \frac{A_t^2}{h}
\right] \psi = 0. \ee
Before we proceed with a discussion of solutions, notice that there is a scaling
symmetry \bea  &z& \ \to \lambda z \ , \ \ x\to \lambda x \ , \ \
Q \to Q /   \lambda \ , \nonumber \\  \ \ &\mu& \ \to \mu
/\lambda \ , \ \ \mathcal{B} \to \mathcal{B} /\lambda^2 \ , \ \ T
\to T/\lambda, \eea so we should only be reporting on
scale-invariant quantities, such as $T/\mu$, $\mathcal{B} / \mu^2$,
$Q/\mu$, etc.
It is convenient to introduce the scale-invariant parameter
\be \beta = \frac{\sqrt{\mathcal{B}}}{q\mu} \ee
to describe the effect of the magnetic field $\mathcal{B}$ of the second $U(1)$.

The system is defined uniquely by specifying the parameters $q$ and $\Delta$. One can then vary the other parameters to study the behavior of the system.
For fixed values of the scale-invariant parameters $Q/\mu$ and $\beta$, we can solve the wave equation \eqref{eqw} and obtain $\mu = \mu_c$ as an eigenvalue. Then from \eqref{eq7}, we deduce the critical temperature at which the second-order phase transition occurs,
\be\label{eqT} \frac{T_c}{\mu} = \frac{T}{\mu_c} =  \frac{3}{4\pi \mu_c} \left[ 1 - \frac{\mu_c^2 (1+ q^4\beta^4\mu_c^2)}{12}
\right]. \ee
For $Q=0$, we recover the homogeneous solution. The maximum critical temperature is obtained for $\beta =0$. In this case, we recover the Reissner-Nordstr\"om black hole. As we increase $\beta$, the temperature \eqref{eqT} decreases. For a given $\beta>0$, the black hole is of the Reissner-Nordstr\"om form with effective chemical potential
\be \mu_{\mathrm{eff}}^2 = \mu_c^2 (1+q^4\beta^4 \mu_c^2)~. \ee
The scalar wave equation is the same as its counterpart in a Reissner-Nordstr\"om background, but with effective charge
\be q_{\mathrm{eff}}^2 = \frac{q^2}{1+q^4\beta^4 \mu_c^2}~, \ee
so that $q_{\mathrm{eff}} \mu_{\mathrm{eff}} = q\mu_c$.

It is known \cite{HartnollPRL101} that the instability \cite{Breitenlohner:1982jf} occurs for all values of $q_{\mathrm{eff}}$, including $q_{\mathrm{eff}}=0$, if $\Delta \le \Delta_\ast$, where $\Delta_\ast =\Delta_+$ for $m^2 = - \frac{3}{2}$, or explicitly,
\be \Delta_\ast = \frac{3+\sqrt{3}}{2} \approx 2.366 ~. \ee
For $\Delta \le \Delta_\ast$, $\beta$ can increase indefinitely. The critical temperature has a minimum value and as $\beta\to\infty$, $T_c$ diverges.

For $\Delta > \Delta_\ast$, $q_{\mathrm{eff}}$ has a minimum value at which the critical temperature vanishes and the black hole attains extremality. This is found by considering the limit of the near horizon region \cite{Horowitz:2009ij,AST}. One obtains
\be\label{eqqmin1} q_{\mathrm{eff}} \ge q_{\mathrm{min}} \ \ , \ \ \ \ q_{\mathrm{min}}^2 = \frac{3+2\Delta (\Delta -3)}{4}. \ee
At the minimum ($T_c=0$), $\mu_{\mathrm{eff}}^2 =12$, and $\beta$ attains its maximum value,
\be\label{eqbmax} \beta \le \beta_{\mathrm{max}} \ \ , \ \ \ \ \beta_{\mathrm{max}}^4 = \frac{1}{12q_{\mathrm{min}}^2} \left( \frac{1}{  q_{\mathrm{min}}^2} -\frac{1}{q^2} \right). \ee
This limit is
reminiscent of the Chandrasekhar and Clogston limit
\cite{instability} in a S/F system, in which a ferromagnet at
$T=0$ cannot remain a superconductor with a uniform condensate.

In the inhomogeneous case ($Q\ne 0$), the above argument still holds with the replacement $m^2 \to m^2 + Q^2$.
The effect of this modification is to increase the minimum effective charge to
\be\label{eqqmin} q_{\mathrm{min}}^2 = \frac{3+2\Delta (\Delta -3)+2Q^2}{4}, \ee
and thus decrease the maximum value of $\beta$ \eqref{eqbmax}. We always obtain a critical temperature which is lower than the corresponding critical temperature (for same $\beta$) in the homogeneous case ($Q=0$).

Notice also that $Q^2\le Q^2_{\mathrm{max}}$, where the maximum value is attained when $q_{\mathrm{min}} = q$ (so that $\beta_{\mathrm{max}} =0$). We deduce from \eqref{eqqmin},
\be\label{eqQmax} Q_{\mathrm{max}}^2 = 2q^2 - \frac{3}{2} - \Delta (\Delta -3). \ee
Now let us add a magnetic interaction term to the action, \be\label{eqV} S_{\mathrm{int}} = \xi \int d^4
x \sqrt{-g} \ |\mathcal{F}^{AB}\partial_B \phi|^2~. \ee
The wave equation is modified to \bea \psi'' + \left[ \frac{h'}{h}
- \frac{2}{z} \right] \psi' - \frac{Q^2}{h} \left[ 1- \xi\mathcal{B}^2 z^4
\right] \psi & & \nonumber\\
- \frac{1}{h} \left[ \frac{m^2}{z^2} - q^2
\frac{A_t^2}{h} \right] \psi &=& 0~. \eea
\begin{figure}[t]
\includegraphics[width=.45\textwidth]{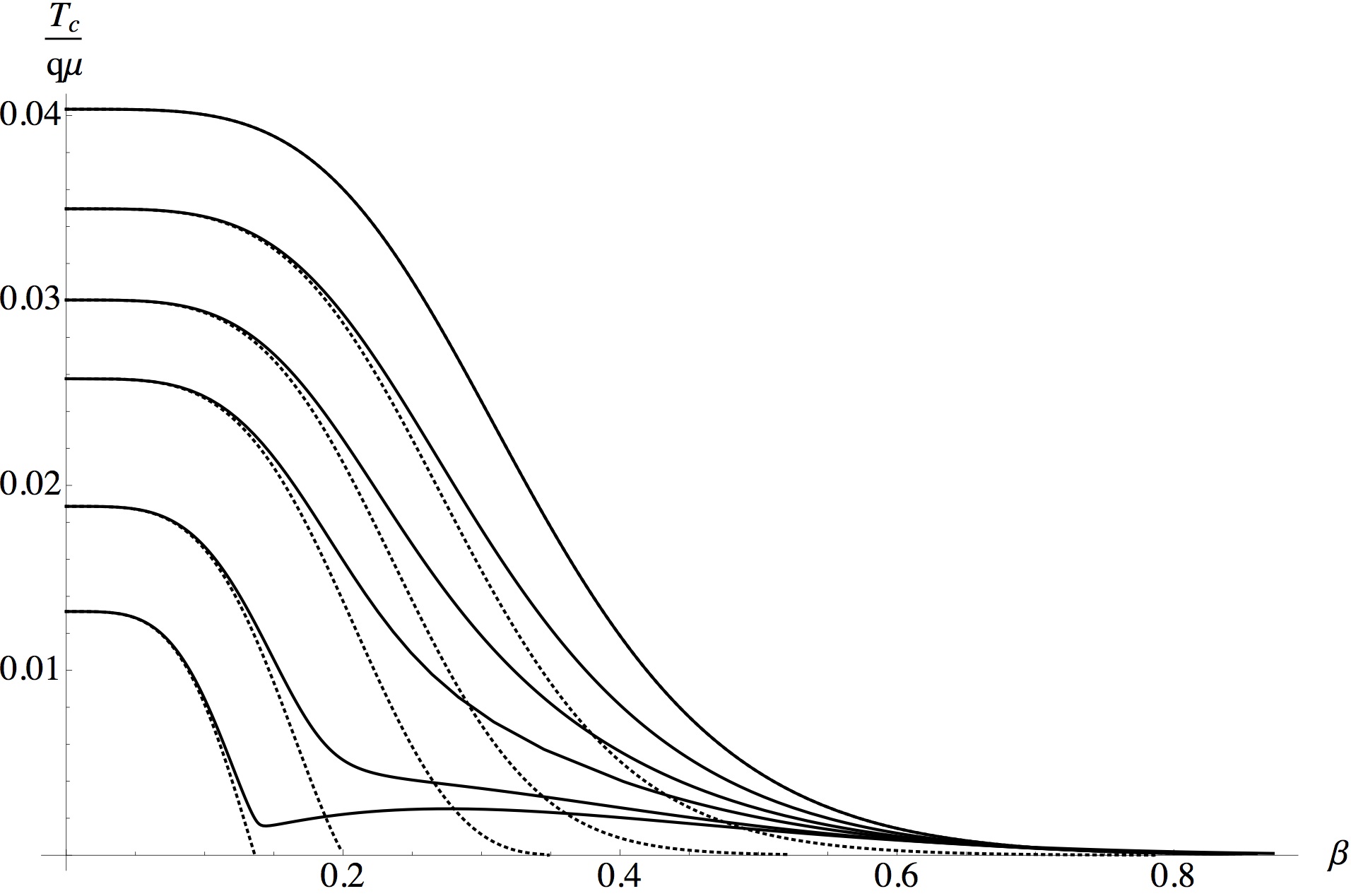}
\caption{The critical temperature $vs.$ the magnetic field numerically calculated with $q=10$ and $\Delta=5/2$.  The dotted lines are calculated with $\xi = 0$ while the solid use $\xi=.10$.  Starting from the top, on the vertical axis, the lines are $\frac{Q^2}{(q\mu)^2} = 0,~.05,~.10,~.15,~.25,$ and $.35$.}
\label{fig1}
\end{figure}
Evidently, if we set $Q=0$, the effect of the interaction term \eqref{eqV} disappears, therefore the homogeneous solution is unaltered. For $Q\ne 0$, we obtain modified solutions. The behavior is shown in figure \ref{fig1}.  The figure also displays the effect of $Q$ on $\beta_{\mathrm{max}}$ \eqref{eqbmax} for $\xi = 0$.

\begin{figure}[t]
\includegraphics[width=.49\textwidth]{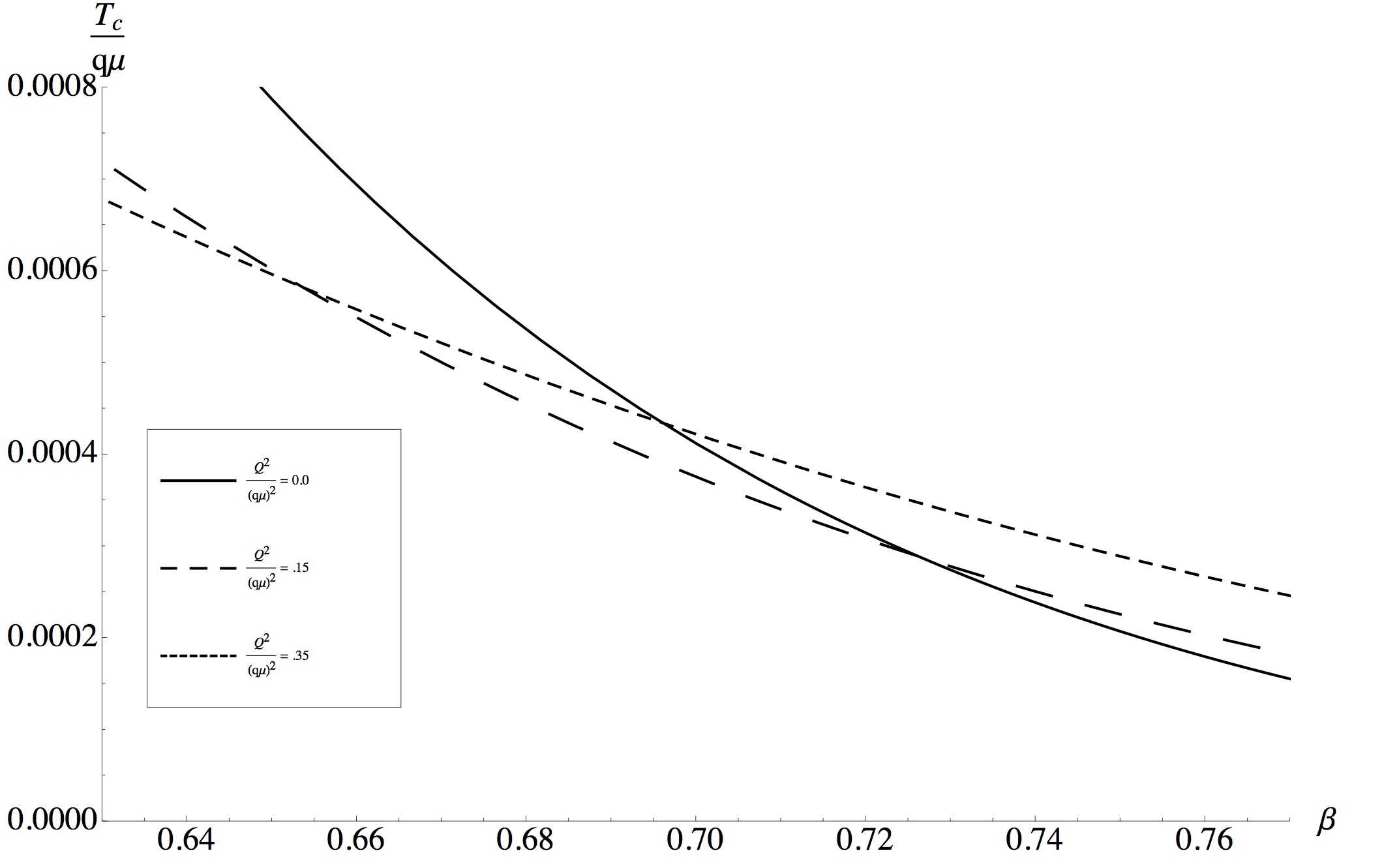}
\caption{The top line on the left-hand side of the graph corresponds to the homogeneous solution, with lines $\frac{Q^2}{(q\mu)^2}=.15, .35$ below.  The critical temperature of the homogeneous solution is found to decrease below the inhomogeneous lines for large $\beta$.  We used $q=10$, $\Delta = 5/2$, and $\xi = .10$.  }
\label{fig2}
\end{figure}

The interaction term \eqref{eqV} alters the near horizon limit of the theory so that $m^2 \to m^2 + Q^2 \left(1-\xi  q^4\beta^4 \mu_c^4\right)$.  The minimum effective charge is found to be dependent on $\xi$, as is $\beta_{\mathrm{max}}$,
\be
\beta_{max}^4 = \frac{\tilde q^2}{12 \tilde q_{min}^2} \left(\frac{\tilde q^2}{\tilde q_{min}^2 } - \frac{1}{q^2} \right),
\ee
where $\tilde q^2 = 1 -\frac{6\xi Q^2}{q^2}$, and $\tilde q_{min}^2 = \frac{3+2\Delta (\Delta -3)+2Q^2\left(1-12\xi\right)}{4}$.
For a range of $\xi$, an increase in $Q$ is seen to increase $\beta_{\mathrm{max}}$.  The modifications are most pronounced for large $\beta$ leading to temperatures which are higher than the critical temperature of the corresponding homogeneous solution.  Figure \ref{fig2} displays the effect of the potential on the homogeneous solution as well as two inhomogeneous cases.

\emph{In conclusion}, we have developed a gravitational dual
theory for the FFLO state of condensed matter.  The gravitational
theory consists of two $U(1)$ gauge fields and a scalar coupled to
a charged AdS black hole. The first gauge field
produces the instability for a condensate to form, while the
second controls the paramagnetic effect. In the absence of an
interaction of the magnetic field with the scalar field, the system
possesses dominant homogeneous solutions for all allowed values of the magnetic field. In the presence of
the interaction term, at low temperatures, the system is shown
to possess a higher critical temperature for a scalar field with
spatial modulation compared to the homogeneous solution.


\emph{Acknowledgments.} This work is dedicated to the memory of
our collaborator in this work, Petros Skamagoulis. We wish to thank
Stefanos Papanikolaou for illuminating discussions.
  J. A. acknowledges support from the Office of Research at
  the University of Michigan-Flint. G. S. is supported by the US Department
of Energy under grant DE-FG05-91ER40627.

\end{document}